\documentclass[10pt,conference]{IEEEtran}

\IEEEoverridecommandlockouts

\AtBeginDocument{%
  \providecommand\BibTeX{{%
    \normalfont B\kern-0.5em{\scshape i\kern-0.25em b}\kern-0.8em\TeX}}}

\newif\ifdraft
\drafttrue
\usepackage[numbers,sort&compress,square]{natbib}
\usepackage{soul}
\usepackage{graphicx}
\usepackage{xurl}
\usepackage{booktabs}
\usepackage{academicons}
\usepackage{multibib}
\usepackage{natbib}
\usepackage{hhline}
\usepackage{xcolor}
\usepackage[compatibility=false]{caption}
\usepackage{graphicx}
\usepackage{colortbl}
\usepackage{multirow}
\usepackage{tabularx}
\usepackage[most]{tcolorbox}
\usepackage{amsmath}
\usepackage{makecell}
\usepackage{fontawesome5}
\usepackage{textgreek}
\usepackage{changepage} 
\usepackage{tabularx}
\usepackage{siunitx}
\usepackage{stfloats, caption}%
\usepackage{tablefootnote}
\newcolumntype{Y}{>{\centering\arraybackslash}X}
\usepackage{bookmark}
\usepackage{textcomp}
\usepackage{xcolor}
\usepackage{microtype}
\usepackage{booktabs}
\usepackage{etoolbox}
\usepackage{enumitem} 
\usepackage{url}
\usepackage{ragged2e}
\usepackage{amsmath}
\usepackage[]{mdframed}

\usepackage[T1]{fontenc}
\usepackage{helvet}
\usepackage{balance}

\usepackage{tikz}
\usetikzlibrary{positioning,fit,backgrounds,arrows.meta,calc}
\usepackage{xcolor}

\usepackage{color}
\definecolor{gray75}{gray}{.25}
\definecolor{gray70}{gray}{.3}
\definecolor{gray60}{gray}{.4}
\definecolor{gray50}{gray}{.5}
\definecolor{gray40}{gray}{.6}
\definecolor{gray30}{gray}{.7}
\definecolor{gray20}{gray}{.8}
\definecolor{gray15}{gray}{.85}
\definecolor{gray10}{gray}{.9}
\definecolor{gray05}{gray}{.95}

\definecolor{apInk}{HTML}{1F2937}
\definecolor{apMute}{HTML}{6B7280}
\definecolor{apRule}{HTML}{9CA3AF}
\definecolor{apS1}{HTML}{F3F4F6}
\definecolor{apS2}{HTML}{EFF6FF}
\definecolor{apS3}{HTML}{DBEAFE}
\definecolor{apS4}{HTML}{BFDBFE}
\definecolor{apAccent}{HTML}{1E40AF}

\definecolor{rqA}{HTML}{D9EAD3} 
\definecolor{rqAdark}{HTML}{38761D}
\definecolor{rqB}{HTML}{FCE5CD} 
\definecolor{rqBdark}{HTML}{B45F06}
\definecolor{rqC}{HTML}{F4CCCC} 
\definecolor{rqCdark}{HTML}{990000}
\definecolor{rqD}{HTML}{CFE2F3} 
\definecolor{rqDdark}{HTML}{0B5394}
\colorlet{sep}{black!55}

\newlength{\apFigW}

\newcommand{\apStage}[3]{%
  \parbox[t]{\dimexpr\apFigW-16pt\relax}{\raggedright
    \textcolor{apAccent}{\scriptsize\bfseries\MakeUppercase{stage #1}}%
    \hspace{5pt}\textcolor{apInk}{\footnotesize\bfseries #2}\\[1.5pt]
    {\footnotesize\color{apInk}#3}}%
}

\definecolor{graytable}{gray}{.95}

\definecolor{headertable}{rgb}{0.0, 0.0, 0.0}

\definecolor{steBoxLine}{rgb}{0.0, 0.0, 0.0}

\definecolor{purple}{HTML}{3a0480}

\mdfdefinestyle{steboxSummaryWhite}{%
linecolor=steBoxLine,linewidth=1pt,%
leftmargin=0cm,rightmargin=0cm,%
roundcorner=5pt,
topline=true,bottomline=true,rightline=false,leftline=false,%
backgroundcolor=gray!15,
frametitlebackgroundcolor=white,
frametitlerule=true,
frametitlerulewidth=1pt
}

\mdfdefinestyle{stebox5}{%
linecolor=steBoxLine,
linewidth=1pt,%
leftmargin=0cm,rightmargin=0cm,%
roundcorner=5pt,
topline=false,bottomline=false,rightline=false,leftline=true,%
backgroundcolor=gray!15
}

\newcommand{\rqsub}[2]{%
  \par\vspace{0.5em}\noindent\hspace*{1em}%
  \begin{minipage}{\dimexpr\linewidth-1em\relax}
    \rqboxSecond{\faAngleDoubleRight\hspace{0.05cm}\textbf{RQ\textsubscript{#1}}: \textit{#2}}
  \end{minipage}\par}

\newcommand{\rqboxSecond}[1]{%
  \colorbox{gray!15}{\parbox{0.97\linewidth}{#1}}%
}

\author{\IEEEauthorblockN{Template}}

\newcolumntype{P}[1]{***{\raggedright\let\newline\\\arraybackslash\hspace{0pt}}m{#1}}

\newcolumntype{C}[1]{***{\centering\let\newline\\\arraybackslash\hspace{0pt}}m{#1}}

\begin{document}

\title{Cognitive Debt in the GenAI Era - Listening to Practitioners’ Voices on the Loss of Understanding\\
}

\author{
\IEEEauthorblockN{%
Theocharis Tavantzis\IEEEauthorrefmark{1},
Bianca Trinkenreich\IEEEauthorrefmark{2},
Michaela Greiler\IEEEauthorrefmark{3},
Júlia Condé Araújo\IEEEauthorrefmark{2},
Marina Condé Araújo\IEEEauthorrefmark{2},\\
Italo Santos\IEEEauthorrefmark{4},
Alexander Serebrenik\IEEEauthorrefmark{5},
Margaret-Anne Storey\IEEEauthorrefmark{6},
Maria Teresa Baldassarre\IEEEauthorrefmark{7}}
\IEEEauthorblockA{%
\IEEEauthorrefmark{1}Aalborg University, Denmark --- thta@cs.aau.dk\\
\IEEEauthorrefmark{2}Colorado State University, USA 
\IEEEauthorrefmark{3}Codalytics, Austria  
\IEEEauthorrefmark{4}University of Hawaii at Manoa, USA \\
\IEEEauthorrefmark{5}Eindhoven University of Technology, The Netherlands
\IEEEauthorrefmark{6}University of Victoria, Canada 
\IEEEauthorrefmark{7}University of Bari, Italy 
}
}










\maketitle
\begin{abstract}
Fast-paced AI-assisted software development risks reducing developers' understanding of the systems they build. This growing gap between what the system does and what developers understand about it, described as ``cognitive debt'', is not new, but AI has exacerbated it. This has led to a recent surge in blog posts and online discussions, in which practitioners recognize and write about cognitive debt, and share their experiences well before academic studies can capture them. To understand their concerns and inform future research and practice, we conduct a grey literature review of 76 sources, focusing on how cognitive debt is conceptualized and manifests, its contributing factors, its consequences, and the practices to prevent, detect, and mitigate it. We find that cognitive debt can accumulate unnoticed behind working code and passing tests, particularly when developers accept AI-generated solutions without building understanding. Practitioners report consequences such as harder maintenance and debugging, fragmented knowledge in the team, and reduced productivity. They propose practices such as documenting design rationale, review-sized changes, comprehension gates, and knowledge-sharing rituals, along with signals like reliability metrics to detect it. Our findings give software teams a structured view of emerging practical approaches in AI-assisted development and stress the need to preserve human understanding alongside faster code production.

\end{abstract}

\section{Introduction}
\label{sec:introduction}

Maintaining software requires developers to understand the systems they change~\cite{von1995program}, the change itself, and why they make it~\cite{storey2006theories}. To build this understanding, developers engage in cognitive processes and form mental representations of a program's structure and behavior~\cite{storey2006theories, fagerholm2022cognition}, which in turn support their decision-making and reasoning about their decisions~\cite{storey2006theories}. However, building such mental models is an effortful and incremental process~\cite{heinonen2023slr}, which does not necessarily follow the fast pace of GenAI-embedded software development.

At the team level, understanding is not necessarily distributed equally among team members. Instead, \emph{shared understanding} refers to the collective availability of knowledge required for the team to perform work tasks. Effective software teams coordinate this expertise by knowing where relevant knowledge resides, where it is needed, and how to utilize it~\cite{faraj2000coordinating}. Thus, a lack of individual understanding is not necessarily a problem, provided that the team has access to the relevant understanding and knowledge when needed. The problem starts when no one else in the team covers gaps in individual knowledge.

This growing gap between what a software system does and what the team understands about it has been described as \emph{cognitive debt}~\cite{storey2026technical}. Cognitive debt is no longer entirely hypothetical: practitioners report adopting AI-generated solutions and code snippets without fully understanding them~\cite{tavantzis2026displacement}. The problem itself is not new; indeed, developers have always reused libraries, frameworks, and copied and pasted parts of code that they did not fully understand \cite{Wang2013}. What GenAI changes is the scale and speed at which this happens. GenAI tools can accelerate the pace at which functionality is added~\cite{peng2023impactaideveloperproductivity}, and early research suggests that teams may produce software faster than they can build and maintain an understanding of it~\cite{miller2026driversOfGenAItools, thoughtworks2026cognitive}. Seen through the lens of sufficiency, the risk is not incomplete understanding per se, but understanding that is insufficient for the software work teams need to perform, potentially hindering extensibility, maintenance, and decision-making.

Nevertheless, cognitive debt remains an emerging and loosely defined concept. The term has only recently gained traction in software engineering~\cite{storey2026cognitive}, and related concerns are often discussed without using the term itself. 
Recognizing cognitive debt as a distinct problem in AI-assisted development and finding ways to manage it are therefore at an early stage, with efforts driven so far mainly by practitioners. Practitioners are sharing their experiences in blog posts, company engineering articles, and online discussions, well before these experiences can be studied and published in peer-reviewed venues. This knowledge is timely but scattered and often uses inconsistent terms. A Grey Literature Review (GLR) surfaces practitioners' experiences, perspectives, and practices that are not yet published in the academic literature, complementing evidence from peer-reviewed studies~\cite{garousi2016need4GreyLiterature, garousi2020benefittingGLR}.

To scope our study, we therefore adopt a working definition of \emph{cognitive debt} as the accumulated gap between the software understanding available to an individual or team and the understanding required to perform current or future software engineering work~\cite{storey2026technical}. This definition served as a guide for identifying relevant accounts, including those that describe such gaps without naming them cognitive debt. How practitioners themselves conceptualize the phenomenon is a question we examine rather than presuppose.

Despite practitioners' growing attention, it remains unclear how AI-induced cognitive debt is understood and manifests in software practice, what contributes to its accumulation, when it becomes consequential, and how to address it. We therefore examine practitioners' perspectives along four dimensions: manifestation, contributing factors, consequences, and management strategies. 
We ask the following research questions (RQs):

\rqboxSecond{
  \faQuestionCircle
  \hspace{0.05cm}
\textbf{RQ\textsubscript{1}}: \textit{How does AI-induced cognitive debt manifest in software engineers' discussions?}}
\vspace{0.5em}

\rqboxSecond{
  \faQuestionCircle
  \hspace{0.05cm}
\textbf{RQ\textsubscript{2}}: \textit{What factors are reported to contribute to the accumulation of AI-induced cognitive debt?}}
\vspace{0.5em}

\rqboxSecond{
  \faQuestionCircle
  \hspace{0.05cm}
\textbf{RQ\textsubscript{3}}: \textit{What consequences of AI-induced cognitive debt are reported for software systems, engineers, and teams?}}
\vspace{0.5em}

\rqboxSecond{
  \faQuestionCircle
  \hspace{0.05cm}
\textbf{RQ\textsubscript{4}}: \textit{What practices or interventions are reported to manage AI-induced cognitive debt across its lifecycle?}}

\begin{adjustwidth}{2em}{0pt}

\rqsub{4.1}{What practices or interventions are reported for \textbf{preventing} AI-induced cognitive debt?}

\rqsub{4.2}{What \textbf{signals and metrics} are reported for identifying cognitive debt that already exists?}

\rqsub{4.3}{What practices are reported for \textbf{mitigating} cognitive debt, once it has been incurred?}

\end{adjustwidth}
\vspace{2mm}
To answer our RQs, we conduct a GLR of 76 sources following established guidelines in software engineering \cite{garousi2019guidelines} and a recent GLR on a similarly new concept \cite{fawzy2026VibeCoding}.
Our findings indicate that practitioners describe cognitive debt as gaps in understanding that surface when code needs to be changed or explained (RQ\textsubscript{1}). They link it to rapid AI code generation, shallower reviews, and delivery pressure (RQ\textsubscript{2}), with consequences for maintenance, developer expertise, and team knowledge (RQ\textsubscript{3}). Practitioners propose smaller changes, comprehension gates, and documented decisions to prevent the debt; reliability metrics and audits to detect it; and knowledge sharing to repay it (RQ\textsubscript{4}). The insights contribute to a better understanding of the potential cognitive consequences of AI adoption and can inform software engineering practices and the design of AI tools that support continued learning, reflection, and critical engagement rather than unreflective reliance.

\section{Background and Related Work}
\label{sec:relatedWork}



GenAI tools can enhance software engineers' efficiency and learning~\cite{ulfsnes2024transformingSD, pereira2025GenAIinSD}, but also introduce risks of \emph{cognitive offloading}, in which individuals delegate cognitive work to external resources to reduce mental effort~\cite{risko2016cognitive,jiang2026coding,brandebusemeyer2026experienceWithGenAI}. While beneficial in the short term, this may weaken developers' engagement with code, learning, and problem-solving abilities~\cite{baltes2026anendlessstreamai,kharrufa2026LLMsInSEprojects,shen2026aiimpactsskillformation,tavantzis2026displacement,shukla2025deskilling}. These concerns highlight a potential gap between what developers can produce with GenAI and what they understand about the resulting systems~\cite{storey2026technical,qiao2026fasterwisergithubcopilot}. Although program comprehension has long involved constructing mental models~\cite{von1995program}, GenAI enables developers to produce functional systems without necessarily developing such understanding.

The notion of \emph{debt} is a widely used metaphor in the SE community to describe cumulative adverse effects. Originating from Cunningham's concept of \emph{technical debt}~\cite{cunningham1993wycash}, the metaphor initially referred to the implied future costs of rework, maintenance, and refactoring resulting from short-term development decisions~\cite{avgeriou2016dagstuhl}. It has since expanded beyond code quality to encompass social and human aspects of software development, including \emph{social debt}, arising
from suboptimal organizational and interpersonal relationships~\cite{tamburri2013social}, \emph{learning debt}, resulting from insufficient investment in developers' learning~\cite{hicks2023coding}, and \emph{human debt}, capturing the accumulated consequences of insufficient psychological safety and inclusive support~\cite{guenes2026humanDebt}. More recently, \emph{competence debt}~\cite{luders2026competenceDebt} and \emph{epistemic debt}~\cite{sankaranarayanan2026mitigatingepistemicdebt} have further extended the metaphor. Within this broader landscape, \emph{comprehension debt} and \emph{cognitive debt} describe the growing gap between software functionality and developers' understanding, individually and collectively~\cite{storey2026technical,ahmad2026comprehensiondebtgenaiassistedsoftware,alakmeh2026comprehensionDebt}.

Focusing on individual-level software understanding, Alakmeh et al.~\cite{alakmeh2026comprehensionDebt} conducted a controlled study to investigate how AI reliance can obscure developers' task comprehension. They showed that while AI-steered participants could complete tasks successfully, they demonstrated significantly lower ability to explain and transfer their knowledge and understanding. These findings empirically support the concern that productivity gains may co-exist with accumulated comprehension debt.

An influential contribution by Storey~\cite{storey2026technical} outlines a triple debt model, which extends technical debt with cognitive and intent debt as complementary dimensions of software health. While technical debt resides in the code, cognitive debt reflects the erosion of the shared understanding and mental models among developers and intent debt captures the loss of externalized goals and rationale that both humans and AI systems need to work effectively within the codebase.

Starr and Storey~\cite{starr2026theorytroubleshooting} offer a complementary contribution by introducing the Theory of Troubleshooting, defined as the cognitive process of understanding and constructing a mental model of the causes of unexpected system behavior. Their findings suggest that the rising cognitive cost of troubleshooting AI-generated code may significantly challenge developers' ability to sustain understanding of a system over its lifetime.

In summary, research on cognitive debt in GenAI-assisted software engineering is growing rapidly, but remains fragmented. Prior studies identify cognitive offloading and the resulting loss of understanding, while more recent works use comprehension and cognitive debt to name this loss at the individual and group levels, respectively. What is missing is a systematic
understanding of how practitioners recognize, experience, and respond to cognitive debt in everyday software development. Building on prior analyses of practitioners' concerns about AI-assisted development, including the erosion of developer competence~\cite{baltes2026anendlessstreamai}, we
synthesize their accounts of how such debt emerges, what consequences they associate with it, and which practices they describe for managing it.


\section{Research Design}
\label{sec:research_design}

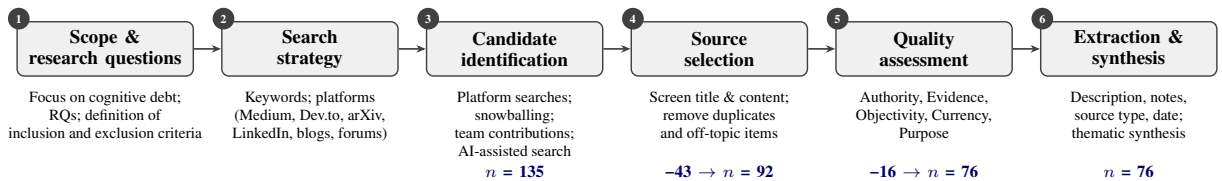
\begin{figure*}[b]
    \centering
    \resizebox{0.9\linewidth}{!}{
        \begin{tikzpicture}[
  stage/.style={draw=black!75, line width=0.6pt, fill=black!6, rounded corners=3pt,
                minimum width=2.6cm, minimum height=0.8cm, inner xsep=2pt,
                align=center, font=\footnotesize\bfseries},
  badge/.style={circle, fill=black!75, text=white, font=\tiny\bfseries,
                minimum size=0.32cm, inner sep=0pt},
  desc/.style={align=center, font=\fontsize{6.5}{7.5}\selectfont},
  count/.style={align=center, font=\scriptsize\bfseries, text=blue!40!black},
  flow/.style={-{Stealth[length=4pt, width=3.2pt]}, line width=0.7pt, draw=black!75},
  node distance=0.4cm
]
  \node[stage] (s1) {Scope \&\\research questions};
  \node[stage, right=of s1] (s2) {Search\\strategy};
  \node[stage, right=of s2] (s3) {Candidate\\identification};
  \node[stage, right=of s3] (s4) {Source\\selection};
  \node[stage, right=of s4] (s5) {Quality\\assessment};
  \node[stage, right=of s5] (s6) {Extraction \&\\synthesis};
 
  \foreach \i in {1,...,6}
    \node[badge] at ([xshift=1pt, yshift=1pt] s\i.north west) {\i};
 
  \foreach \i/\j in {1/2, 2/3, 3/4, 4/5, 5/6}
    \draw[flow] (s\i.east) -- (s\j.west);
 
  \node[desc, below=0.1cm of s1] (d1) {Focus on cognitive debt;\\RQs; definition of\\inclusion and exclusion criteria};
  \node[desc, below=0.1cm of s2] (d2) {Keywords; platforms\\(Medium, Dev.to, arXiv,\\LinkedIn, blogs, forums)};
  \node[desc, below=0.1cm of s3] (d3) {Platform searches;\\snowballing;\\team contributions;\\AI-assisted search};
  \node[desc, below=0.1cm of s4] (d4) {Screen title \& content;\\remove duplicates\\and off-topic items};
  \node[desc, below=0.1cm of s5] (d5) {Authority, Evidence,\\Objectivity, Currency,\\Purpose};
  \node[desc, below=0.1cm of s6] (d6) {Description, notes,\\source type, date;\\thematic synthesis};
 
  \node[count, below=0.1cm of d4] (c4) {\textminus43 $\rightarrow$ $n$ = 92};
  \node[count, anchor=north] (c3) at (c4.north -| s3) {$n$ = 135};
  \node[count, anchor=north] (c5) at (c4.north -| s5) {\textminus16 $\rightarrow$ $n$ = 76};
  \node[count, anchor=north] (c6) at (c4.north -| s6) {$n$ = 76};
\end{tikzpicture}
    }
    \caption{Grey Literature Review process}
    \label{fig:method}
\end{figure*}

We conducted a systematic GLR to investigate how AI-induced cognitive debt is described, experienced, and addressed in software engineering practice. We followed the guidelines of Garousi et al.~\cite{garousi2019guidelines} and a recent GLR on vibe coding by Fawzy et al.~\cite{fawzy2026VibeCoding}, which applied them to a similarly newly introduced concept. Figure~\ref{fig:method} outlines our six-stage process, from scoping and search to quality evaluation and synthesis. 


\subsection{Search Strategy}
\label{sec:search}

Searching grey literature differs from searching academic databases because web sources are inconsistently indexed, search-engine rankings are dynamic, and relevant practitioner terminology may vary substantially. Following established recommendations for GLRs~\cite{garousi2019guidelines}, we defined the search engine and platforms, search strings, search depth, source types, and stopping criteria before conducting the final search.

\subsubsection{\textbf{Search Term Development}}
\label{sec:search-terms}

Because terminology in grey literature is not standardized~\cite{garousi2019guidelines}, we developed the search terms iteratively using terminology associated with cognitive debt (e.g., cognitive debt, comprehension debt, and epistemic debt) and AI in software engineering (e.g., AI-generated code, agentic AI coding, and vibe coding).
We then examined the results this combination of the concept groups triggered. This step showed that some terms we had considered as relevant are used by practitioners for distinct phenomena. For example, \emph{knowledge debt} mostly refers to missing documentation or knowledge silos independent of AI~\cite{bloomfire2026knowledge}. In this regard, we iteratively revised the search string based on our working definition until it retrieved relevant sources with minimal noise. The final search string was:

\begin{center}
\begin{tcolorbox}[
  colback=white,
  colframe=black,
  boxrule=0.8pt,
  arc=0pt,            
  outer arc=0pt,
  left=6pt, right=6pt, top=6pt, bottom=6pt,
  width=0.95\linewidth
]
(``cognitive debt'' OR ``loss of understanding'' OR ``lack of understanding'' OR ``understanding gap'') AND (``software engineering'' OR ``software development'') AND (``generative AI'' OR ``GenAI'' OR ``large language model'' OR ``LLM'' OR ``AI coding assistant'' OR ``AI-generated code'' OR ``AI-assisted development'' OR ``vibe coding'' OR ``agentic AI coding'')
\end{tcolorbox}
\end{center}

We validated our final search string against a quasi-gold standard~\cite{fawzy2026VibeCoding} of sources we had already identified as foundational, including Storey's blogs (GR9, GR10), Osmani’s comprehension-debt piece (GR12), the Thoughtworks Technology Radar entry (GR4), and Fowler's fragments (GR39, GR54). Our final string retrieved them.

\subsubsection{\textbf{Search Strategy and Documentation}}
\label{sec:web-search}

We used Google as the primary search engine for the GLR. Practitioners' accounts are broadly distributed, and Google can retrieve diverse grey literature across blogs, forums, and reports~\cite{tahir2023multivocalReview, adams2016GreyLiteraturePublicHealth, adams2016SearchingGreyLiterature}. However, searching on a general-purpose search engine may introduce bias towards well-indexed pages. To increase search coverage, we complemented the general-purpose engine with platform-specific searches, backward snowballing \cite{wohlin2022}, AI-assisted searching, and crowdsourced additions, where authors added relevant sources the searches had not retrieved. We applied the same selection and quality assessment to these sources.

Regarding the AI-assisted searching, we used Claude Fable 5.1 model, following general principles of using LLMs in software engineering research~\cite{baltes2026guidelinesempiricalstudiessoftware}. The search happened in September 2026, and the prompt included the working definition, sibling terminology, target platforms, and the already-included sources as an exclusion list. Then, one author screened the candidate sources against the same criteria as the human-derived ones. The prompt is in our Figshare repository \href{https://figshare.com/s/d11f06d8c16ecf5db18d}{\faDatabase}.

Regarding targeted sources, Garousi et al.~\cite{garousi2019guidelines} recommend identifying and complementing general-purpose search engines with specialized websites and forums. Our prior knowledge and early searches showed that discussions around cognitive debt focus on a small number of platforms, including personal blogs of well-known software engineers, practitioners' publishing platforms (e.g., LinkedIn, Medium, Dev.to), discussion forums (Reddit), and arXiv as a preprint server and source of grey literature~\cite{garousi2019guidelines}.
We selected these platforms because they host sources that we were already aware of and considered foundational. For each website, we used internal search, where feasible, with the same or adapted search terms.

As web-search results change over time, we recorded the complete search procedure so that our review process can be audited and traced. For each retrieved result, we recorded the platform, source title, author or organization, URL, and publication date. For LinkedIn posts, we extracted the date from the post timestamp. We conducted the searches between May and September 2026. \textbf{The complete sources log is included in our Figshare replication package} \href{https://figshare.com/s/d11f06d8c16ecf5db18d}{\faDatabase}.


\subsection{Source Selection}
\label{sec:selection}

We conducted our screening procedure in three stages: (1) removal of duplicates and near-duplicates, (2) screening of titles, and (3) full-source screening against the inclusion and exclusion criteria (see Table~\ref{tab:criteria}). Since grey-literature sources rarely contain abstracts, initial relevance was assessed from the title, search-result snippet, author, and visible source content.

Two pairs of authors independently screened the sources assigned to them. Initially, the pairs assessed a pilot sample of 10 sources to calibrate the interpretation of the inclusion and exclusion criteria. For example, this pilot sharpened the boundary between a substantive discussion of cognitive debt (IC6) and a superficial contribution (EC4). We discussed unclear cases in pairs, and a third researcher joined when we could not reach agreement. We used these discussions to sharpen how we interpreted and operationalized the criteria in practice.

\begin{table}[b]
\centering
\caption{Inclusion/exclusion criteria used to select sources.}
\label{tab:criteria}
\rowcolors{1}{white}{gray!15}
\small
\begin{tabular}{p{0.05\linewidth}p{0.85\linewidth}}
\hline
\textbf{ID} & \textbf{Criterion} \\
\hline
IC1 &
Sources written in English \\

IC2 &
Sources published within the defined review period, from early 2025 (emergence of agentic AI) to September 2026 \\

IC3 &
Sources publicly available in full text \\

IC4 &
Sources relevant to at least one of the research questions \\

IC5 &
Sources situated in a software engineering context \\

IC6 &
Sources that discuss cognitive debt providing reflections, evidence, or examples, rather than merely mentioning it \\

\hline

EC1 &
Duplicate or mirrored sources \\

EC2 &
Sources focused exclusively on technical aspects, without discussing cognitive or human consequences \\

EC3 &
Undated, anonymous, or otherwise unverifiable sources \\

EC4 &
Short-form content with insufficient substantive information, evidence, or reflections \\

\hline
\end{tabular}
\end{table}

\subsection{Quality Assessment}
\label{sec:quality}

Because grey literature lacks uniform peer review, it requires careful assessment beyond inclusion and exclusion criteria~\cite{garousi2019guidelines,kamei2021gl1se}. Thus, we scored each source against five dimensions from Garousi et al.'s quality checklist~\cite{garousi2019guidelines}, in the form used by Fawzy et al.~\cite{fawzy2026VibeCoding}. Each dimension was scored from 0 (absent) to 2 (fully satisfied), giving a maximum of 10.

\begin{itemize}[leftmargin=*]
    \item \textbf{Authority}: Is the author or organization identifiable, and is there evidence of relevant software engineering expertise or experience?

    \item \textbf{Evidence}: Does the source explain the experience, observations, data, examples, or reasoning supporting its claims?

    \item \textbf{Objectivity}: Does the source distinguish evidence or experience from opinion and avoid primarily promotional claims?

    \item \textbf{Currency}: Is the source published within the review period?

    \item \textbf{Purpose}: Is the source intended to inform, reflect, or explain rather than to promote or market?
\end{itemize}

We omitted the impact and outlet-control criteria~~\cite {garousi2019guidelines} because impact metrics are not comparable across the platforms we searched and would have systematically favored LinkedIn content. We used the midpoint as a threshold; thus, we retained sources scoring at least 5. The quality assessment followed the same pairing structure as the Source Selection (Section~\ref{sec:selection}). We resolved disagreements through synchronous or asynchronous communication, and a third author independently scored 20\% of the sources to increase robustness.



\subsection{Data Extraction}
\label{sec:extraction}

To enhance the traceability of our processes and decisions, we extracted data in a common spreadsheet. For each source, we included the title, author, platform, publication date, URL, a short description of the source's argument, its quality score, and analytic notes. We also extracted behavioral units from each source. A \emph{behavioral unit} is a single, descriptive statement that captures information relevant to the RQs. In the context of this study, that could include (a) a manifestation of cognitive debt, (b) a factor or practice that accumulates it, (c) a consequence of cognitive debt, or (d) a practice that mitigates or identifies cognitive debt. Each source contributes several behavioral units, related to varying RQs (\textit{see our replication package} \href{https://figshare.com/s/d11f06d8c16ecf5db18d}{\faDatabase}).

\vspace{-1mm}
\subsection{Qualitative Analysis}
\label{sec:analysis}

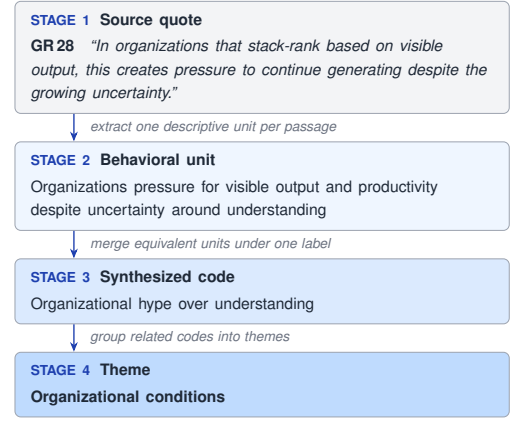
\begin{figure}[t]
  \centering
  \resizebox{0.75\linewidth}{!}{
  \begin{tikzpicture}[
    font=\sffamily,
    stage/.style={draw=apRule, line width=0.5pt, rounded corners=3pt,
                  inner xsep=8pt, inner ysep=6pt,
                  text width=\dimexpr\apFigW-16pt\relax, align=left},
    flow/.style={-{Stealth[length=5pt,width=4pt]}, line width=0.7pt, draw=apAccent},
    op/.style={font=\scriptsize\sffamily\itshape, text=apMute, anchor=west,
               inner sep=2pt, text width=5.2cm, align=left},
  ]

  \node[stage, fill=apS1] (A) at (0,0) {\apStage{1}{Source quote}{%
    \textbf{GR\,28}\quad\emph{``In organizations that stack-rank based on visible output,
    this creates pressure to continue generating despite the growing uncertainty.''}}};

  \node[stage, fill=apS2, below=14pt of A] (B) {\apStage{2}{Behavioral unit}{%
    Organizations pressure for visible output and productivity despite uncertainty around understanding}};

  \node[stage, fill=apS3, below=14pt of B] (C) {\apStage{3}{Synthesized code}{%
    Organizational hype over understanding}};

  \node[stage, fill=apS4, below=14pt of C] (D) {\apStage{4}{Theme}{%
    \textbf{Organizational conditions}}};

  \coordinate (ax) at ($(A.south west)!0.12!(A.south east)$);
  \draw[flow] (ax |- A.south) -- (ax |- B.north)
    node[op, midway, xshift=6pt] {extract one descriptive unit per passage};
  \draw[flow] (ax |- B.south) -- (ax |- C.north)
    node[op, midway, xshift=6pt] {merge equivalent units under one label};
  \draw[flow] (ax |- C.south) -- (ax |- D.north)
    node[op, midway, xshift=6pt] {group related codes into themes};

  \end{tikzpicture}}
  \caption{Example of Data Analysis}
  \vspace{-5.3mm}
  \label{fig:analysis-process}
\end{figure}

We analyzed the behavioral units using Reflexive Thematic Analysis~\cite{braun2006thematicAnalysis, braun2019rta}. We began by familiarizing ourselves with the sources and extracted behavioral units to identify recurring patterns. We then generated candidate codes for each unit and grouped conceptually related codes into potential themes. We subsequently reviewed the themes in relation to the full set of behavioral units and our RQs to identify overlaps, inconsistencies and potential refinements. Finally, we defined and named the final themes and developed descriptions of their meaning, which informed the synthesis and reporting of our analysis. Figure~\ref{fig:analysis-process} illustrates a worked example, moving from source quotes to themes.

The first author extracted behavioral units and preliminary codes from the first 31 sources, producing an initial codebook that the authors discussed in a joint meeting and refined asynchronously. Then, six authors extracted and coded behavioral units from the remaining sources using the codebook, adding new codes where needed. Two authors cross-checked the extracted units and codes, resolving any disagreements with the respective coder through discussion. The first author then grouped the codes into candidate themes, which were discussed in a collective meeting and refined asynchronously. Throughout, the process was interpretive and reflexive rather than one in which independent coders are expected to converge on a single “correct” interpretation. In line with Braun and Clarke's recommendation~\cite{braun2019rta}, we therefore did not calculate inter-coder agreement. Instead, we ensured quality through collaborative coding, cross-checking, and documented discussions. 

\subsection{Evidence Strength and Triangulation}
\label{sec:triangulation}

Practitioner sources vary substantially in the strength of evidence they provide, spanning from a single anecdote to reflections across teams and organizations. We therefore considered a theme more strongly supported when it (1) appeared across multiple independent sources, (2) was backed by concrete examples or data, (3) appeared in high-scoring sources, and (4) was not attributable solely to repeated or syndicated content. The last criterion matters in our context, because many accounts are responses to Storey's post~\cite{storey2026cognitive}. Notably, sources that just reframed her position without adding reflections were not included in the final analysis.







\section{Results}
\label{sec:results}

\begin{figure*}[b]
  \centering
  \resizebox{0.95\linewidth}{!}{%
\begin{tikzpicture}[
  font=\sffamily,
  card/.style={inner sep=0.14cm, anchor=north west, align=left, text=black, execute at begin node={\hyphenpenalty=10000 \exhyphenpenalty=10000 }, font=\sffamily\fontsize{6.2}{7.4}\selectfont},
  rqtext/.style={text=white, align=center, inner sep=1pt, font=\sffamily\bfseries\fontsize{5.8}{6.8}\selectfont},
]
\coordinate (origin) at (0,0);
\coordinate (row0) at (1.57,0);
\node[card, text width=8.40cm] (c00) at ([xshift=0.00cm]row0) {{\bfseries\fontsize{7.2}{7.4}\selectfont Competing interpretations of cognitive debt}\\[2pt]Bad engineering practice that AI amplifies {\color{sep}\textbullet} Distinct from technical and legacy debt {\color{sep}\textbullet} Misdiagnosed as technical debt {\color{sep}\textbullet} Debt-metaphor framings {\color{sep}\textbullet} Verification debt {\color{sep}\textbullet} Intent debt {\color{sep}\textbullet} Epistemic / theory debt {\color{sep}\textbullet} Comprehension debt};
\node[card, text width=6.50cm] (c01) at ([xshift=8.80cm]row0) {{\bfseries\fontsize{7.2}{7.4}\selectfont Erosion of collective human understanding}\\[2pt]Inability to explain own or merged code {\color{sep}\textbullet} Mental-model gap {\color{sep}\textbullet} Understanding as the bottleneck {\color{sep}\textbullet} Loss of code comprehension through AI-assisted implementation};
\node[card, text width=4.95cm] (c02) at ([xshift=15.70cm]row0) {{\bfseries\fontsize{7.2}{7.4}\selectfont Hidden nature of cognitive debt}\\[2pt]Silent accumulation {\color{sep}\textbullet} AI-induced overconfidence};
\node[fit=(c00)(c01)(c02), inner sep=0] (band0) {};
\begin{scope}[on background layer]
  \draw[fill=rqA, draw=black!65, line width=0.4pt, rounded corners=2pt] (c00.north west |- band0.north) rectangle (c00.south east |- band0.south);
  \draw[fill=rqA, draw=black!65, line width=0.4pt, rounded corners=2pt] (c01.north west |- band0.north) rectangle (c01.south east |- band0.south);
  \draw[fill=rqA, draw=black!65, line width=0.4pt, rounded corners=2pt] (c02.north west |- band0.north) rectangle (c02.south east |- band0.south);
  \fill[rqAdark, rounded corners=1.5pt] (0,0 |- band0.north) rectangle (1.45, 0 |- band0.south);
\end{scope}
\node[rqtext] at (0.72, 0 |- band0.center) {RQ1\\Manifestation};
\path ([yshift=-0.18cm]band0.south west) coordinate (row1);
\node[card, text width=3.81cm] (c10) at ([xshift=0.00cm]row1) {{\bfseries\fontsize{7.2}{7.4}\selectfont Individual behavior}\\[2pt]Uncritical acceptance of AI output {\color{sep}\textbullet} Cognitive surrender {\color{sep}\textbullet} Confirmation bias {\color{sep}\textbullet} Bypassing understand-while-coding};
\node[card, text width=4.36cm] (c11) at ([xshift=4.21cm]row1) {{\bfseries\fontsize{7.2}{7.4}\selectfont Properties of the tools}\\[2pt]Generation speed and volume outpace comprehension {\color{sep}\textbullet} Agents lack tacit and contextual team knowledge {\color{sep}\textbullet} Ambiguity of natural language specification};
\node[card, text width=3.39cm] (c12) at ([xshift=8.97cm]row1) {{\bfseries\fontsize{7.2}{7.4}\selectfont Teams' processes}\\[2pt]Weakened code review as knowledge channel {\color{sep}\textbullet} Rationale not captured};
\node[card, text width=3.82cm] (c13) at ([xshift=12.76cm]row1) {{\bfseries\fontsize{7.2}{7.4}\selectfont Organizational conditions}\\[2pt]Organizational hype over understanding {\color{sep}\textbullet} Velocity incentives and deadline pressure};
\node[card, text width=3.67cm] (c14) at ([xshift=16.98cm]row1) {{\bfseries\fontsize{7.2}{7.4}\selectfont Interdependencies with other debt}\\[2pt]Reinforcing loop with technical debt and over-complex code};
\node[fit=(c10)(c11)(c12)(c13)(c14), inner sep=0] (band1) {};
\begin{scope}[on background layer]
  \draw[fill=rqB, draw=black!65, line width=0.4pt, rounded corners=2pt] (c10.north west |- band1.north) rectangle (c10.south east |- band1.south);
  \draw[fill=rqB, draw=black!65, line width=0.4pt, rounded corners=2pt] (c11.north west |- band1.north) rectangle (c11.south east |- band1.south);
  \draw[fill=rqB, draw=black!65, line width=0.4pt, rounded corners=2pt] (c12.north west |- band1.north) rectangle (c12.south east |- band1.south);
  \draw[fill=rqB, draw=black!65, line width=0.4pt, rounded corners=2pt] (c13.north west |- band1.north) rectangle (c13.south east |- band1.south);
  \draw[fill=rqB, draw=black!65, line width=0.4pt, rounded corners=2pt] (c14.north west |- band1.north) rectangle (c14.south east |- band1.south);
  \fill[rqBdark, rounded corners=1.5pt] (0,0 |- band1.north) rectangle (1.45, 0 |- band1.south);
\end{scope}
\node[rqtext] at (0.72, 0 |- band1.center) {RQ2\\Contributing\\factors};
\path ([yshift=-0.18cm]band1.south west) coordinate (row2);
\node[card, text width=5.58cm] (c20) at ([xshift=0.00cm]row2) {{\bfseries\fontsize{7.2}{7.4}\selectfont For the software system}\\[2pt]Harder to modify and debug generated code {\color{sep}\textbullet} Increased fragility: regressions, incidents, tipping point {\color{sep}\textbullet} Degradation of system quality attributes};
\node[card, text width=6.56cm] (c21) at ([xshift=5.98cm]row2) {{\bfseries\fontsize{7.2}{7.4}\selectfont For the engineer}\\[2pt]Hesitation and loss of confidence {\color{sep}\textbullet} Anxiety, fatigue, burnout {\color{sep}\textbullet} Skill atrophy and deskilling {\color{sep}\textbullet} Reduced ability to steer and verify AI {\color{sep}\textbullet} Loss of identity and ownership {\color{sep}\textbullet} Productivity decrease};
\node[card, text width=7.70cm] (c22) at ([xshift=12.95cm]row2) {{\bfseries\fontsize{7.2}{7.4}\selectfont For the team and organization}\\[2pt]Cognitive meltdown (no human understands the system) {\color{sep}\textbullet} Knowledge fragmentation and concentration {\color{sep}\textbullet} Dependency on AI vendors and tools {\color{sep}\textbullet} Slower onboarding {\color{sep}\textbullet} Erosion of trust and accountability};
\node[fit=(c20)(c21)(c22), inner sep=0] (band2) {};
\begin{scope}[on background layer]
  \draw[fill=rqC, draw=black!65, line width=0.4pt, rounded corners=2pt] (c20.north west |- band2.north) rectangle (c20.south east |- band2.south);
  \draw[fill=rqC, draw=black!65, line width=0.4pt, rounded corners=2pt] (c21.north west |- band2.north) rectangle (c21.south east |- band2.south);
  \draw[fill=rqC, draw=black!65, line width=0.4pt, rounded corners=2pt] (c22.north west |- band2.north) rectangle (c22.south east |- band2.south);
  \fill[rqCdark, rounded corners=1.5pt] (0,0 |- band2.north) rectangle (1.45, 0 |- band2.south);
\end{scope}
\node[rqtext] at (0.72, 0 |- band2.center) {RQ3\\Consequences};
\path ([yshift=-0.18cm]band2.south west) coordinate (row3);
\node[card, text width=6.35cm] (c30) at ([xshift=0.00cm]row3) {{\bfseries\fontsize{7.2}{7.4}\selectfont Prevention}\\[2pt]Comprehension gate {\color{sep}\textbullet} Human engagement {\color{sep}\textbullet} Scope to low-risk work {\color{sep}\textbullet} Review-sized changes {\color{sep}\textbullet} Externalize intent {\color{sep}\textbullet} Executable guardrails};
\node[card, text width=6.76cm] (c31) at ([xshift=6.75cm]row3) {{\bfseries\fontsize{7.2}{7.4}\selectfont Signals and metrics}\\[2pt]Reliability and process metrics as indicators {\color{sep}\textbullet} Cognitive-debt scoring and audits {\color{sep}\textbullet} Behavioral signals of cognitive debt};
\node[card, text width=6.74cm] (c32) at ([xshift=13.91cm]row3) {{\bfseries\fontsize{7.2}{7.4}\selectfont Mitigation and repayment}\\[2pt]Code review {\color{sep}\textbullet} Pairing and mentoring {\color{sep}\textbullet} Share knowledge {\color{sep}\textbullet} Scheduled transfer {\color{sep}\textbullet} AI-assisted understanding {\color{sep}\textbullet} Practices repurposed};
\node[fit=(c30)(c31)(c32), inner sep=0] (band3) {};
\begin{scope}[on background layer]
  \draw[fill=rqD, draw=black!65, line width=0.4pt, rounded corners=2pt] (c30.north west |- band3.north) rectangle (c30.south east |- band3.south);
  \draw[fill=rqD, draw=black!65, line width=0.4pt, rounded corners=2pt] (c31.north west |- band3.north) rectangle (c31.south east |- band3.south);
  \draw[fill=rqD, draw=black!65, line width=0.4pt, rounded corners=2pt] (c32.north west |- band3.north) rectangle (c32.south east |- band3.south);
  \fill[rqDdark, rounded corners=1.5pt] (0,0 |- band3.north) rectangle (1.45, 0 |- band3.south);
\end{scope}
\node[rqtext] at (0.72, 0 |- band3.center) {RQ4\\Management};
\end{tikzpicture}
  }
  \caption{Overview of themes and codes}
  \label{fig:themes}
\end{figure*}
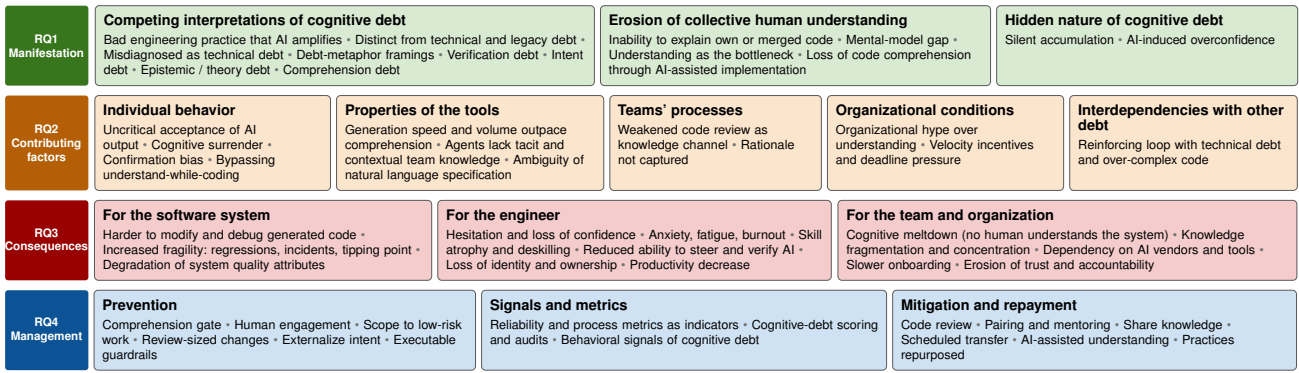

In this section, we describe the grey-literature corpus (Section~\ref{sec:corpus}) and then present the themes identified for each RQ (Figure~\ref{fig:themes}). We follow the four RQs to report how practitioners describe the manifestations of cognitive debt (RQ\textsubscript{1}), the factors that contribute to its accumulation (RQ\textsubscript{2}), its consequences for software systems, engineers, and teams (RQ\textsubscript{3}), and the practices they report for managing (preventing, detecting, and mitigating) it (RQ\textsubscript{4}). Throughout this section, we use \textbf{bold} for themes and \textsc{small caps} for codes, and we refer to sources by their identifiers (e.g., GR1), listed in our replication package. Since our goal is to characterize practitioners' accounts rather than estimate their prevalence, we do not report code frequencies.

\vspace{-1mm}
\subsection{Overview of the Grey-Literature Corpus}
\label{sec:corpus}

Our identification strategies yielded 135 candidate sources in total, combining search-engine and platform searches, backward snowballing, crowdsourced additions from the authors’ saved material, and AI-assisted search. Of these, 92 passed the screening phase, and 76 passed the quality assessment, forming the final corpus.

The corpus comprises 11 company or vendor articles, one industry report, 58 practitioner articles, one technical document, and five arXiv preprints. No forum threads met the inclusion and quality criteria. Sources were published between February 2025 and September 2026, with the volume peaking in March 2026 (16 sources), the month after Storey’s post~\cite{storey2026cognitive} brought the term into wider circulation. Sources are referred to as GR1–GR76 throughout.

\subsection{RQ\textsubscript{1}: Manifestation}
\label{sec:rq1}
\label{sec:rq1}


We identified \textbf{Competing Interpretations of Cognitive Debt} in discussions about whether AI creates a new type of debt, how it differs from technical debt, and how its costs are understood. GR16 described cognitive debt as \textsc{a bad engineering practice that AI amplifies}, arguing that \textit{``cognitive debt is not an AI problem. It is a human problem that AI amplifies''}, a view also shared by GR22. GR5 agreed that the problem is not new, but emphasized \textit{``the scale of the problem being created''}. The \textsc{distinction from technical and legacy debt} appears in GR40's differentiation between debt in software artifacts and missing human understanding, while GR13 considered missing design-level knowledge more consequential than missing implementation details. This distinction is not always recognized, as cognitive debt may be \textsc{misdiagnosed as technical debt}, with engineers blaming \textit{``technical debt: messy code, poor architecture, hurried implementations''} (GR8) rather than missing understanding. We also identified \textsc{debt-metaphor framings} that describe the trade-off between immediate productivity and future costs, such as \textit{``trading current productivity for future capability''} (GR24).

Authors used related terms to emphasize different aspects of missing understanding. \textsc{Verification debt} is observed when you are  \textit{``approving diffs you haven't fully read''} (GR38). The \textsc{intent debt (loss of rationale) } refers to missing knowledge about why decisions were made, as \textit{``the codebase persists. The justification does not''} (GR46). \textsc{Epistemic/theory debt} focuses on gaps in developers' mental models when cognitive work is outsourced to AI, a concept discussed in GR41 through another study. Finally, \textsc{comprehension debt} describes the individual gap between existing code and what the developer understands (GR22), a symptom not of cognitive decline but of \textit{``our habits and processes being broken''} (GR76), its team-level counterpart is the next theme. 

The \textbf{Erosion of Collective Human Understanding} describes a growing mismatch between a software system and what a team of developers knows about it. GR2 describes this as \textit{``the growing gap between what [the] codebase does and what [the] team actually comprehends''}. One manifestation is the \textsc{inability to explain own or merged code}, when developers cannot \textit{``explain how [the] function works, [even if] it passes all the tests''} (GR51). Another is the \textsc{mental-model gap}, when developers' understanding does not match the system's actual behavior (GR55), or they miss \textit{``the nuanced reasoning that went into the implementation''} (GR44). AI-assisted development can widen these gaps by making code generation faster than comprehension. This creates \textsc{understanding as the bottleneck}, described as the \textit{``gap between output velocity and comprehension velocity''} (GR28). One explanation is the \textsc{loss of code comprehension through AI-assisted implementation}, as developers may bypass the learning through \textit{``understand-while-coding''} (GR1), missing opportunities to build mental models of the implementation (GR27).

The \textbf{Hidden Nature of Cognitive Debt} describes how missing understanding may go unnoticed until developers need it (e.g., during a change or an incident) (GR18). A \textsc{silent accumulation} happens when \textit{``the code works, the tests pass, the features ship''} (GR28), even though developers lack understanding. Polished code may also be mistaken for understood code (GR18). The \textsc{AI-induced overconfidence} adds to this problem, as AI may \textit{``invent a confident-sounding reason, which is worse than admitting it doesn't know''} (GR7).

\subsection{RQ\textsubscript{2}: Contributing Factors}
\label{sec:rq2}


\textbf{Individual Behavior} captures how software engineers work with AI tools. Several sources mentioned the \textsc{uncritical acceptance of AI-generated output}. GR5 wondered \textit{“how much vibe coded software is out there in the wild that just appears to work?”}. This is also discussed as a \textsc{cognitive surrender}, or \textit{“the quiet habit of accepting the machine's output without doing the harder thinking yourself”} (GR18). A \textsc{confirmation bias} arises when polished AI outputs lower scrutiny: as GR58 put it \textit{“When I open a PR that was written with AI assistance, [...] The code looks more polished.[...] My lizard brain whispers: 'This seems fine.'”}. Developers may also \textsc{bypass understand-while-coding} and miss opportunities to build understanding when AI handles the implementation. GR6 describes a growing distance between developers acting as ``orchestrators'' and the code being committed, while GR45 argues that removing the effort of writing code also removes the learning that comes with it.

\textbf{Properties of the Tools} covers capabilities and weaknesses of AI tools that contribute to cognitive debt. On the capability side, \textsc{generation speed and volume outpace comprehension}, as GR31 illustrates \textit{“AI produce[s] code at a speed that outruns the team's ability to build a mental model”}. 
On the other hand, as a limitation, \textsc{agents lack tacit and contextual team knowledge}, which erodes the reasoning and assumptions behind implementation choices.
GR16 shared a narrative to support this view, saying that \textit{“When a senior engineer leaves today, their understanding survives in memory, notes, commit messages, chat history [...]. When an AI coding session ends, the knowledge that lived in that session's context is simply gone. No forwarding address.”} Another limitation of the tools is the \textsc{ambiguity of natural language specification,} as \textit{``what [we] say is often not what [we] mean, and LLMs fill in ambiguity with assumptions (or hallucinations), which leads to: more review, more agent revisions, more tokens burned, and more disconnection from what is being created.”} (GR6).

\textbf{Teams' Processes} can contribute to cognitive debt when AI-generated code changes how developers review and share knowledge. The growing volume of AI-generated code can \textsc{weakened code review as knowledge channel}. GR72 describes how code review has shifted from \textit{``a conversation between developers''} to validating generated code. As GR28 explains, \textit{``the volume of generated code exceeds the bandwidth available for deep review. Something has to give, and typically it is review depth''}, leaving less room for discussing and understanding implementation decisions. Besides, with \textsc{rationale not captured}, important context hides behind past decisions and may remain only in developers' heads (GR61).

\textbf{Organizational conditions} can also favor code production over understanding. A growing \textsc{organizational hype over understanding} happens when no strategy mitigates cognitive debt, and developers are under \textit{``pressure to continue generating despite the growing uncertainty''} (GR28). GR17 similarly links \textsc{velocity incentives and deadline pressure} to the temptation to accept AI-generated solutions without examining them: \textit{``accepting AI-generated code is effortless. Understanding it takes work. Deadline pressure always favors speed''}.


Finally, \textbf{interdependencies with other debt} captures how existing technical debt and unnecessary code complexity may contribute to the accumulation of cognitive debt. GR4 described a \textsc{reinforcing loop with technical debt and over-complex code} that \textit{``makes systems progressively harder to reason about''}. As GR5 noted, \textit{“once the overly complex code is committed, it is much harder to recognize the complexity is not needed”}.


\subsection{RQ\textsubscript{3}: Consequences}
\label{sec:rq3}


\textbf{Consequences for the Software System} can happen when cognitive debt affects software maintenance, reliability, and quality. Practitioners described \textsc{increased effort to modify and debug generated code}, when developers \textit{“hit a tricky bug and have to spend time building up the mental model to figure out what's wrong”} (GR5), offsetting the time saved during generation. As this lack of understanding accumulates, systems become \textsc{increasingly fragile}, where \textit{“small changes trigger unexpected failures, fixes introduce regressions and cleanup efforts increase risk instead of reducing it”} (GR4). These consequences also affect incident resolution, turning what would have been \textit{“a ten-minute fix when someone understood the system”} into \textit{“a four-hour forensic investigation when no one does”} (GR28). Beyond maintainability, practitioners raised concerns about \textsc{degradation of system quality attributes}, particularly security, when \textit{“developers lack the internal understanding or landscape knowledge to respond to incidents or perform effective threat modeling”} (GR34).

The \textbf{Consequences for the Engineer} capture how cognitive debt affects developers' confidence, emotional well-being, skills, ability to supervise AI, professional identity, and productivity. Practitioners described \textsc{hesitation and loss of confidence}, as cognitive debt \textit{“show[s] up as hesitation [.. and] engineers avoid parts of the system they no longer trust”} (GR18), limiting their ability to modify code confidently. These challenges also contribute to \textsc{anxiety, fatigue and burnout}, as \textit{“an anxiety [they] didn't have in the understand-while-coding era.”} (GR1). Beyond these immediate consequences, \textsc{skill atrophy and deskilling} emerge when developers lose opportunities to learn, since \textit{“without the friction and challenges that come with working with code directly, their ability to learn is seriously diminished.”} (GR6). This erosion of understanding also leads to a \textsc{reduced ability to steer and verify AI}, as \textit{“effectively using Claude requires supervision, and supervising Claude requires the very coding skills that may atrophy from AI overuse.”} (GR6). GR38 described the \textsc{loss of professional identity and ownership}, noting what a developer once said: \textit{``I used to be a craftsman... and now I feel like I am a factory manager''}. Finally, \textsc{Productivity decrease} was mentioned when faster code generation does not translate into faster delivery, as the time saved writing code may be spent reviewing and understanding it: \textit{“Writing time ↓ Review time ↑ Lead time... doesn't change”} (GR72).

Regarding \textbf{Consequences for the Team and Organization}, practitioners warned about a \textsc{cognitive meltdown}, or \textit{“a state where no human is able to understand how a system works given any reasonable amount of time”} (GR36). They also described \textsc{knowledge fragmentation and concentration}, as \textit{“the workload concentrates because a small group handles deep system context, security-sensitive areas, and ownership boundaries.”} (GR60), while others warned about \textit{“a risk of a kind of organizational atomization”} (GR34). Another consequence is \textsc{dependency on AI vendors and tools}, where \textit{“what used to be a skill that they could execute with just a keyboard and text editor suddenly required a subscription to an AI model provider.”} (GR6). These knowledge gaps also contribute to \textsc{slower onboarding}, particularly when new engineers cannot rely on colleagues to explain implementation decisions (GR14).
Beyond knowledge transfer, cognitive debt contributes to \textsc{erosion of trust and accountability}, challenging the assumption that developers understand the code they contribute, since \textit{“normally you trust that the author understands their own code. With AI, that assumption is no longer safe.”} (GR14).

\subsection{RQ\textsubscript{4}: Management 
}
\label{sec:rq4}

We report practitioners’ accounts along the three sub-questions: prevention (RQ\textsubscript{4.1}), signals and metrics for recognition (RQ\textsubscript{4.2}), and mitigation once incurred (RQ\textsubscript{4.3}).

\subsubsection{\textbf{Prevention}}

\textsc{Comprehension Gate} practices require developers to demonstrate understanding of AI-generated code before accepting or merging it, so they \textit{``never ship code [they] can't explain.''} (GR2). GR8 suggests requiring at least one person to fully understand each AI-generated change before deploying it to production. Practitioners also emphasize the importance of understanding code beyond whether it works. GR62 argues that \textit{``the author can explain the happy path, the failure modes, and why this approach was chosen''}. Some propose making comprehension checks an explicit part of the development workflow, such as an agent configuration that quizzes developers about generated code and prevents commits until they can explain it (GR10).

\textsc{Human Engagement} emphasizes actively involving developers in implementation rather than delegating the entire process to AI. GR24 advocates using LLMs as \textit{``pairing partners rather than replacements''}, while GR14 describes an approach in which \textit{``you write the function signatures, types, and interfaces. AI fills the implementation''}. Other practitioners recommend developing an initial solution or expectation before consulting AI (GR22), retaining human responsibility for architectural decisions (GR27), or regularly coding without AI assistance to maintain programming skills (GR24). As GR18 puts it, developers should let AI handle repetitive work \textit{``while you stay mentally engaged with the structure.''}

\textsc{Scope to Low-Risk Work} captures recommendations to limit AI delegation depending on the criticality and longevity of the code. Practitioners suggest 
\textit{``preferring AI for scaffolding and refactoring, not architectural decisions''} (GR25),
using AI for boilerplate, test generation, and repetitive transformations (GR24), while retaining responsibility for consequential design decisions. GR14 identifies authentication, payments, data deletion, and database migrations as areas requiring particularly strong human ownership. Others distinguish disposable prototypes from durable software, arguing that short-lived code can tolerate shallower understanding than systems requiring long-term maintenance (GR34).

\textsc{Review-Sized Changes} addresses the volume of AI-generated code by limiting changes to what humans can meaningfully review. GR6 describes the personal rule as \textit{``[they] never generate more than [they] can review in a sitting.''} Similarly, GR22 argues that \textit{``the unit of review is the unit of comprehension''}, recommending small PRs rather than large generated changes. Practitioners also recommend splitting work into smaller increments and verifying each against specifications before proceeding (GR61). Smaller, atomic changes are additionally presented as easier to understand and revert independently (GR43).

\textsc{Externalizing Intent} focuses on preserving the reasoning and decisions behind code in explicit artifacts. GR7 recommends \textit{``writing the spec for the intent, not the implementation''}, capturing goals, constraints, and an explicit definition of done. Practitioners propose architecture decision records (ADRs), design documents, and repository-level context files to preserve knowledge that might otherwise remain implicit (GR7, GR10). GR25 recommends documenting what a change solves, why an approach was chosen, known edge cases, and considered trade-offs before merging. Others advocate lightweight approaches, including PR descriptions, acceptance criteria, or decisions extracted from agent conversations (GR31, GR63). As per GR15, \textit{``Document the why, not the what.''}

\textsc{Executable Guardrails} describes the use of specifications and automated checks to verify AI-generated implementations. GR1 explains that \textit{``[we] don't need to trust the LLM's output''}, because it can be checked against the specification.
Other approaches include architectural fitness functions to continuously check constraints (GR4) and agent context files that communicate architectural preferences, conventions, and forbidden patterns (GR70).

\subsubsection{\textbf{Signals and Metrics}}




The \textsc{reliability and process metrics as indicators} code captures how metrics such as the number of commits per pull request (GR23) and Mean Time to Recovery can signal cognitive debt.  As GR28 noted:  
\textit{
``The debt is not truly invisible. It eventually appears in reliability metrics: Mean Time to Recovery stretches longer, Change Failure Rate creeps upward.''} 
GR31 further adds:
\textit{``How often does a team revert code because nobody understood the edge case?''}

The \textsc{cognitive-debt scoring and audits} code captures how practitioners consider using special purpose metrics for cognitive debt diagnosis as well as practices for auditing its presence and magnitude over time.
GR19 in their ``Triple Debt Scanner'' post mentions how hard it is for a new engineer to build a mental model as a measurement of cognitive debt, further noting that one must consider both technical and cognitive debt together: 
\textit{``Blast zone detection via heat map: The most dangerous combination in a codebase is high technical debt paired with low cognitive coverage modules that are broken AND that nobody fully understands. The scanner visualizes this as a live heat map and flags these as blast zones: the places where an AI agent, a new hire, or an urgent hotfix is most likely to cause a cascading failure.''} GR27 also mentions that \textit{``None of these metrics measure human understanding directly. They measure the infrastructure for understanding.''}
Practitioners propose not only metrics for measuring and diagnosing cognitive debt, but also how these or similar metrics can be used in periodic audits of comprehension debt (as cognitive debt may grow invisibly over time). 
GR2 suggests: \textit{``Set a calendar reminder for the quarterly cognitive debt audit''} and 
\textit{``identify high-churn files, cross-reference with AI origin, test whether two engineers can explain each one.''}
The \textsc{Behavioral signals of cognitive debt} code captures proposed measures or more subjective signals of cognitive debt based on observed behaviors.  
GR76 describes ``overproduction pressure'' as one such signal. 
GR41 describes a number of behavior signals: 
\textit{``Your pull request reviews are getting shallower. You can confirm that tests pass, but you engage less with the logic.
You can build new features but struggle to maintain old code. New functionality comes easily; debugging a six-month-old function is inexplicably hard.
You paste error messages directly into chat without forming a hypothesis first. The diagnostic step - `what could cause this?' - gets skipped entirely.
Your mental model of your own system is fuzzy. If someone asked you to whiteboard the architecture, the lines get uncertain quickly.
You accept AI-generated solutions without being able to articulate why they work. The answer compiles and passes tests, and that is the entirety of your validation.''}




\subsubsection{\textbf{Mitigation and Repayment}}



\textsc{Code Review} extends beyond defect detection to support collective understanding and knowledge transfer. GR43 argues that \textit{``The primary outcome of code review is knowledge transfer''}, emphasizing its role in exposing developers to unfamiliar parts of a system. Practitioners recommend examining implementation decisions alongside correctness (GR55), rotating reviewers to promote collective ownership and learning (GR58, GR64), and automating mechanical checks while focusing human attention on intent, architecture, and functional correctness (GR57). 

\textsc{Pairing and Mentoring} emphasizes collaborative engagement with AI-generated code to rebuild understanding. Practitioners recommend pair programming alongside refactoring and testing (GR8), including having developers jointly retype and discuss complex generated algorithms (GR44). Others suggest pairing developers with AI agents while maintaining active human involvement. GR17 describes this as \textit{``Pair programming mode -- Agent suggests, you type. You build the mental model through muscle memory.''} These practices emphasize learning through implementation and discussion rather than merely reviewing generated outputs.

The \textsc{Share Knowledge} code focuses on distributing system understanding through rotating code ownership (GR25, GR37), walkthroughs (GR48), and conversations with experienced team members (GR45). GR25 recommends rotating ownership to \textit{``Make sure everyone on the team can describe the system's most critical components.''} Similarly, GR45 emphasizes recovering knowledge from those familiar with a system's history: \textit{``We talk to whoever is still around: the engineers who maintained it, the people who supported it.''} GR63 further suggests sharing knowledge earlier through collaborative discussions of design intent and acceptance criteria.

\textsc{Scheduled Transfer} involves allocating dedicated time to rebuild and share understanding after AI-assisted development. Practitioners recommend sprint-level comprehension reviews of AI-heavy changes (GR2, GR14), regular code walkthroughs and architecture reviews (GR15), and \textit{``understanding-rebuilding sessions''} noted by GR42. These activities help identify poorly understood code and gaps in shared knowledge (GR38). GR36 further emphasizes the need to explicitly plan for cognitive debt repayment, describing it as \textit{``a human rebuilding the theory''} of the system.

The \textsc{AI-assisted Understanding} code captures the use of AI to rebuild and challenge developers' understanding of code. Practitioners suggest using LLMs to explain unfamiliar code and implementation decisions (GR7, GR29, GR50), generate video walkthroughs (GR3), or quiz developers (GR10). GR17 recommends a \textit{``Learning mode - Ask the agent to explain what it's doing. Read the explanation. Don't just accept the code.''} Others advocate adversarial interactions, with GR22 suggesting \textit{``Ask the model to argue against itself''} and GR63 proposing agents that challenge developers' decisions and reasoning. However, GR31 cautions that \textit{``Generated explanations are useful as a starting point. They are not a substitute for the team deciding what it believes.''}

Finally, \textsc{Practices Repurposed} describes using established software engineering activities, including code reading, debugging, testing, and refactoring, to rebuild understanding of AI-generated code (GR35, GR38). GR10 recommends refactoring sessions that \textit{``force deep re-engagement with the code''}, arguing that \textit{``You can't improve something you don't understand.''} Similarly, GR35 emphasizes tracing execution and restructuring generated code to strengthen comprehension. Other approaches include dedicated sessions to revisit AI-generated implementations (GR38) and test-driven development to support verification and understanding (GR8, GR29).


\section{Discussion}
\label{sec:discussion}
\vspace{-1mm}

\subsection{Key observations}

\textbf{Dual role of cognitive surrender.} In RQ\textsubscript{2} (Sec.\ref{sec:rq2}), \textsc{cognitive surrender} appears as a contributing factor to cognitive debt. GR18 describes it as \textit{``the quiet habit of accepting the machine's output without doing the harder thinking yourself''}, while GR22 distinguishes between using AI \textit{``to think or instead of thinking''}. This aligns with previous research showing that people sometimes accept AI-generated answers even when they are incorrect \cite{shaw2026thinking} or do not meet quality attributes~\cite{tavantzis2026displacement}. However, our findings suggest that cognitive surrender may also result from accumulated cognitive debt. RQ\textsubscript{1} (Sec.~ \ref{sec:rq1}) showed gaps in developers' mental models, while RQ\textsubscript{3} (Sec.~ \ref{sec:rq3}) showed a \textsc{reduced ability to steer and verify AI}. GR41 described how accepting unreviewed AI-generated features left the author \textit{``getting lost in [their] own projects with no firm mental model''}. Developers who no longer understand their systems may become increasingly dependent on AI for decisions they cannot independently evaluate. This feedback loop remains to be investigated.

\begin{itemize}
    \item [\faHandPointRight] \textbf{Recommendation for practitioners:} \textit{Treat every AI-accepted change as a claim you need to defend. Before merging the PR, state in your own words what the change does and why this approach was chosen.}
\end{itemize}

\textbf{Approaches to prevent cognitive debt.} Practitioners propose two approaches to prevent cognitive debt. The first requires developers to manually ensure they understand the system through staying engaged during reviewing or coding, while the second emphasises capturing the understanding in artifacts and making them part of the automated workflow. In some cases, this entails mandatory requirements within the AI-assisted development workflow, such as comprehension gates, PR-size limits, documented decisions, executable tests, and agent constraints. Repayment and reconstructing understanding, on the other hand, rely chiefly on traditional engineering activities, such as pair-programming~(GR8, GR17), code review~(GR8, GR43), and walkthroughs~(GR42, GR48).

\begin{itemize}
    \item [\faHandPointRight] \textbf{Recommendation for Team Leads:} \textit{Incorporate comprehension gates as a natural part of the workflow, for example by adding thresholds to PR sizes so that changes remain reviewable and requiring decision documents for non-trivial AI changes. Schedule and protect familiar knowledge-sharing rituals, such as pair-programming and code reviews.}
\end{itemize}

\textbf{Maintaining understanding beyond working code.} Storey~\cite{storey2026technical} distinguished cognitive debt, the erosion of shared understanding, from intent debt, the loss of design rationale. Our RQ\textsubscript{1} findings (Sec.~\ref{sec:rq1}) show how these gaps appear in practitioners' discussions, including developers unable to explain merged code, incomplete mental models, and knowledge concentrated in a single team member. Working code and passing tests, therefore, do not guarantee that others will be able to maintain it. Previous research argued that AI is enabling developers to work in unfamiliar domains, sometimes without the expertise needed to verify their solutions~\cite{baolin2026developer}. For practitioners, the goal is not to avoid AI or understand every line of generated code, but to preserve the required knowledge for future maintenance and to share it beyond one person.

\begin{itemize}
    \item [\faHandPointRight] \textbf{Recommendation for practitioners:} \textit{For any AI-assisted change, record why it was made and which assumptions it relied on, and share that knowledge beyond the developer who supervised the AI through walkthroughs and pairing.}
\end{itemize}

\textbf{Producing code without developing understanding.} In the GenAI era, code production and learning during implementation can be fragmented (Sec.~\ref{sec:rq1}). Developers may produce working solutions without the reasoning that builds a mental model of the system. Routine GenAI use can make users feel they no longer need to understand what they produce \cite{choudhuri2026thinking} and let developers bypass the ``productive struggle'' involved in building expertise, which makes complex problems harder to debut later \cite{baolin2026developer}. Practitioners describe AI as both a contributor to cognitive debt and a means of repaying it. However, GR31 raises an important concern: a plausible AI-generated explanation does not necessarily produce durable human understanding. Just reading the output may not yield the same insights that doing the work would have.

\begin{itemize}
    \item [\faHandPointRight] \textbf{Recommendation for practitioners:} \textit{Use AI to explore alternatives, question design decisions, and explain unfamiliar code, rather than simply accepting generated solutions. For critical systems preserve your mental model by modifying and debugging the code and writing the tests.}
\end{itemize}

\textbf{Recognizing cognitive debt.} While code works, tests pass, and features ship, cognitive debt can remain unnoticed (Sec~\ref{sec:rq1}). This raises a practical question: how can teams identify missing understanding before it affects maintenance? Prior research has shown that code review involves constructing mental models of proposed changes and their relationship to the existing system \cite{gonccalves2025code}. However, confusion during code review is common and can increase developers' effort \cite{ebert2021exploratory}. Moreover, an empirical study found that 42\% of review comments contained suggestions without explanations \cite{widyasari2025explaining}. Code reviews therefore detect cognitive debt only if reviewers probe reasoning and assumptions, not just code behavior.


\begin{itemize}
    \item [\faHandPointRight] \textbf{Recommendation for practitioners:} \textit{Use code review as a comprehension gate. Ask the author to explain the reasoning and the assumptions behind the change. Treat potential lack of explanation as a knowledge gap to close before merging, not as a reason to reject the PR.}
\end{itemize}

\textbf{Comprehension scores should not be used to measure individual performance.}
Our RQ\textsubscript{4} findings include proposals to assess cognitive debt through comprehension scores and regular audits on how well team members understand a feature's purpose, design decisions, and possible modifications. However, GR75 emphasized that \textit{``individual scores should not be tracked or reported.”} Given the organizational pressure for visible productivity identified in RQ\textsubscript{2}, using these scores to evaluate developers could discourage them from acknowledging gaps in their understanding.

\begin{itemize}
    \item [\faHandPointRight] \textbf{Recommendation for organizations and Team Leads:} \textit{If you measure comprehension scores, measure them at the team level and use them only as an indication for knowledge sharing and restructuring. Do not associate them with individual performance.}
\end{itemize}



\vspace{-2mm}
\subsection{Implications for research}

Our findings offer foundations for future work. RQ\textsubscript{1} distinguishes \textsc{intent debt}, the loss of design rationale, from gaps in developers' understanding. However, RQ\textsubscript{2} suggests that these problems may be related, as practitioners identify \textsc{rationale not captured} and \textsc{agents lack tacit and contextual team knowledge} as contributing factors to cognitive debt. The maintenance difficulties and fragmented team knowledge reported in RQ\textsubscript{3} raise the question of whether missing design rationale makes these problems worse. Storey's triple-debt model also proposes that intent debt can lead to cognitive debt when missing rationale prevents developers from building an adequate understanding of the system \cite{storey2026technical}. Our findings provide further motivation to study this relationship empirically and test whether access to intent improves the ability of developers to understand AI-generated software.

In RQ\textsubscript{4}, practitioners propose \textsc{externalizing intent} through design records and PR descriptions, and GR45 suggests using the proportion of system rules whose rationale remains unknown as an indicator of cognitive debt. Future work could test whether this measure reflects developers' actual understanding and predicts their ability to modify code. A related question was also raised for AI agents, which must reconstruct project context from persistent artifacts when previous session context is lost (GR54). Future research could investigate whether access to design rationale improves the ability of both developers and agents to understand and modify software when maintenance tasks pass between them. Finally, the dual role of cognitive surrender suggested by our findings is not established or systematically studied. A longitudinal study could test this effect and whether the recommended practices can rebuild the required understanding to break this loop.


\vspace{-1mm}
\section{Threats to Validity}
\label{sec:threats}
\vspace{-1mm}

In this section, we discuss the threats to the validity of our findings.

\paragraph{\textbf{Construct Validity}} Cognitive debt is a recently introduced term in the AI discussions, so sources may describe it using different terms or use the term for different phenomena. To mitigate the risk of wrongly including or excluding borderline sources, we adopted a working definition, applied explicit inclusion and exclusion criteria, and assessed sources against predefined quality criteria.

\paragraph{\textbf{External Validity}} Our findings rely on grey literature published on blogs and practitioner platforms, whose authors may over-represent specific profiles, such as AI enthusiasts or skeptics. Similarly, they do not capture domain- and role-specific experiences. In this regard, our findings should be interpreted as patterns in practitioners' accounts, rather than evidence or tendencies among the population, which is also why we do not report code frequencies. Therefore, their transferability may depend on contextual factors, such as the type of AI tools and practitioners' experiences with them.

\paragraph{\textbf{Internal Validity}} Reliance on Google may bias retrieval toward well-indexed pages. To mitigate this threat, we constructed the search string systematically and complemented the results with a platform-specific searches, backward snowballing, systematic AI-assisted search and crowd-sourced additions. To limit selection and quality bias, we applied inclusion and exclusion criteria and two authors evaluated each source against five quality criteria proposed by Garousi et al.~\cite{garousi2019guidelines}. On top of that, two authors, not involved in the quality evaluation, cross-checked the sources independently. However, assessment remains a subjective process and the final scores may reflect our biases. Finally, Reflexive Thematic Analysis is by design an interpretative method, which introduces a risk of interpretation bias. We followed Braun and Clarke's methodology~\cite{braun2006thematicAnalysis} and two authors independently cross-checked the extracted themes and codes, resolving disagreements through reflexive discussion; candidate themes were discussed and refined by all authors. 

\section{Conclusion}
\label{sec:conclusion}
\vspace{-1mm}




We conducted a GLR of 76 sources to examine how practitioners understand, experience, and manage cognitive debt. Our findings characterize cognitive debt as a growing gap between software functionality and developers' understanding, with reported consequences for maintainability, developer expertise, and shared team knowledge. Practitioners propose preventive, diagnostic, and mitigation practices emphasizing human engagement, preserving design rationale, and rebuilding understanding. Moreover, AI emerges as both a contributor to cognitive debt and a potential means of addressing it. Future research should empirically investigate how cognitive debt accumulates over time, develop and validate measures of individual and team understanding, and evaluate the effectiveness of proposed mitigation practices in real-world settings. Overall, our findings highlight the importance of considering not only the productivity gains of AI-assisted development, but also whether engineers and teams retain the understanding needed to maintain, evolve, and take responsibility for the software they build.

\section*{Acknowledgments}

This work was initiated at the Copenhagen Symposium, organized by Daniel Russo in May 2026.


\balance
{\small
    \bibliographystyle{IEEEtran}
    \bibliography{references}
}
\end{document}